\documentclass[runningheads]{llncs}
\usepackage[T1]{fontenc}
\usepackage{graphicx}
\begin{document}
\title{Do Gains from Generative AI–Enabled Adaptive Pretesting Persist? Evidence from a Retention Study}
\titlerunning{Do Gains from Generative AI–Enabled Adaptive
Pretesting Persist?}
%
\author{Mahir Akgun\inst{1}\orcidID{0000-0001-6884-0119} \and
Sacip Toker\inst{2}\orcidID{0000-0003-1437-6642}}
\authorrunning{M. Akgun and S. Toker}
%
\institute{College of Information Sciences and Technology, The Pennsylvania State University, University Park, PA 16802, USA \and
Information Systems Engineering, Atilim University, Ankara, TR \\
\email{makgun@psu.edu, \ sacip.toker@atilim.edu.tr}}
%
\maketitle 
\begin{abstract}
Pretesting — attempting problems before instruction — supports learning by activating prior knowledge and sharpening attention to subsequent instruction. Recent work suggests that adaptive AI-assisted pretesting can yield further advantages, particularly for tasks requiring higher-order reasoning, yet it remains unclear whether these gains persist over time. This study examines the durability of learning gains following GenAI-enabled adaptive pretesting over a seven-week retention period. Undergraduate participants completed an adaptive AI-assisted pretesting session, received instruction, and took a baseline assessment, then were randomly assigned to adaptive spaced retrieval practice, fixed spaced retrieval practice, or learner-directed AI-supported study. Multivariate analyses revealed a significant effect of condition on posttest performance and observed practice effort, with retrieval-based conditions outperforming learner-directed study. Findings indicate that adaptive pretesting can elevate initial understanding, but sustained learning depends on how subsequent AI-supported practice is structured.

\end{abstract}

\section{Introduction}
Pretesting involves prompting learners to attempt answers before formal instruction. Although learners typically perform poorly on such initial attempts, a substantial body of research demonstrates that pretesting enhances subsequent learning by activating prior knowledge, revealing knowledge gaps, and sharpening attention to critical information encountered later \cite{carpenter2005application,kornell2009unsuccessful,pan2023prequestioning,pressley1990happens,richland2009pretesting}. These benefits often occur even when initial responses are incorrect \cite{richland2009pretesting}, suggesting that pretesting supports learning through mechanisms beyond simple accuracy.

The role of pretesting is especially salient when learners have immediate access to external information. Ready availability of answers through search engines and intelligent systems can reduce effortful reasoning \cite{sparrow2011google}, encouraging cognitive outsourcing that undermines durable learning. Research on “think-before-search” paradigms shows that generating answers before consulting external resources yields superior recall compared with immediate lookup \cite{giebl2023thinking,giebl2021answer}, underscoring the importance of preserving productive struggle. 

Generative AI (GenAI) introduces new possibilities for adaptive pretesting. Unlike fixed sequences—where learners may provide superficial responses and still advance—GenAI can probe unclear reasoning, request elaboration, and withhold progress until a substantive attempt is made. Emerging evidence suggests that adaptive AI-supported pretesting enhances learning and engagement, particularly for tasks requiring conceptual integration or iterative reasoning \cite{akgun2025struggle}. 

Yet the durability of these gains remains an open question. Learning science has long distinguished short-term performance improvements from long-term retention, emphasizing that gains observed immediately after instruction may dissipate without continued practice  \cite{Ebbinghaus1913Memory}. Two well-established forms of continued practice — spacing and retrieval — offer well-established mechanisms for sustaining learning: spaced practice reliably outperforms massed practice 
\cite{Carpenter2022TheSO}, retrieval practice produces more durable learning than restudying  \cite{roediger2006test}, and their combination yields especially robust outcomes \cite{Carpenter2022TheSO}. This raises the possibility that adaptive pretesting functions as a front-loaded catalyst that prepares learners for subsequent activities rather than as a stand-alone solution.

This study examines two outcome variables: end-of-semester posttest performance, as an indicator of long-term retention and near-transfer, and observed practice effort, derived from AI interaction logs, as a behavioral indicator of productive engagement during the retention phase. Together, these outcomes capture both what students learned and how they engaged with AI-supported practice over time. Two research questions guide the study:

–	RQ1: Do learning gains from GenAI-enabled adaptive pretesting persist over a seven-week retention period when different forms of follow-up practice are provided (adaptive spaced retrieval, fixed spaced retrieval, or learner-directed AI study)?

–	RQ2: Does practice structure affect the quality of learner engagement, as measured by observed practice effort, during AI-supported practice sessions?

\section{Method}
\subsection{Participants}
Participants were undergraduate students enrolled in a large, upper-division course at a public research university in the United States. The course was an applied statistics course designed for students majoring in Cybersecurity or Security, Risk, and Analysis, with the target content for this study focusing on multiple linear regression — a topic that requires both conceptual understanding and applied analytical reasoning. A total of 89 students (after exclusions due to incomplete data or noncompliance with study procedures) were included in the final analyses. Participation occurred as part of regular course activities, and all procedures were approved by the institution's ethics review board.
\subsection{Study Design and Procedure}
The study employed a between-subjects experimental design with three post-baseline conditions following a common initial learning phase. 
\subsubsection{Adaptive AI Agent.}
The adaptive AI agent was implemented using a generative large language model configured via a structured system prompt defining its instructional role, response scope, and escalation logic. During pretesting, the agent applied response-contingent prompting: if a learner’s response revealed a clear misconception, the agent issued a targeted probe; if a response was superficial or incomplete, the agent requested elaboration before proceeding; and if a response demonstrated adequate conceptual engagement, the agent advanced to the next item. Direct solutions and posttest-relevant answers were explicitly excluded from the agent’s allowed outputs during pretesting. In the adaptive retrieval condition (G1), the agent additionally reviewed prior session performance signals across sessions, increasing conceptual depth for content areas with weaker prior responses while maintaining challenge in stronger areas.

\subsubsection{Initial phase.}
During adaptive AI-assisted pretesting, learners attempted domain-relevant questions before receiving explanations, with the AI agent adjusting follow-up questions dynamically. Direct solutions were withheld to preserve productive struggle. After pretesting, all participants received the same instructional materials via standard course resources, followed immediately by a baseline assessment.

\subsubsection{Experimental Conditions.}
All participants completed the same adaptive AI-assisted pretesting phase prior to randomization. Participants were randomly assigned to one of the three conditions that differed in how AI-supported practice was structured over seven weeks:
\begin{itemize}
    \item \textbf{Adaptive Spaced Retrieval Practice (G1):} Participants received spaced retrieval prompts via the AI agent, \textit{adaptively generated based on prior responses and performance signals}.
    \item \textbf{–	Fixed Spaced Retrieval Practice (G2):} Participants engaged in spaced retrieval practice via the same AI agent \textit{using a fixed question sequence}. The agent presented questions one at a time without feedback, adaptation, or response-contingent adjustment.
    \item \textbf{Learner-Directed AI Study  (G3):} Participants interacted with the AI agent in spaced sessions without enforced retrieval. Learners freely directed the interaction, asking questions about the target content; the agent responded without enforcing retrieval attempts, question sequencing, or adaptivity.
\end{itemize}
The number and timing of practice sessions were held constant across all three conditions. Conditions differed intentionally in how AI interaction was structured within sessions: G1 received adaptively generated retrieval prompts with response-contingent feedback; G2 received the same fixed question sequence without adaptation; and G3 engaged in open-ended, learner-directed interaction without enforced retrieval. These differences in interaction structure constitute the experimental manipulation rather than an uncontrolled source of variance. Analyses of submitted session logs indicated that interaction volume — measured by total exchange length per session — did not differ significantly across conditions, supporting the comparability of practice exposure independent of its structure. At the end of the semester, all participants completed the same posttest assessing retention and near-transfer of target concepts.



\subsection{Measures }
\subsubsection{Baseline and Posttest.}
Both the baseline and posttest were parallel forms of a 14-item multiple-choice instrument developed for this study. Each item presented a scenario-based problem situated in a cybersecurity context, requiring students to apply multiple linear regression concepts — including predictor selection, multicollinearity detection, model comparison using adjusted R², and interpretation of regression coefficients. Items were designed to assess both retention of core concepts and near-transfer to novel applied problems. The two forms were counterbalanced such that one served as the baseline and the other as the posttest, with both scored on a 0–100 scale.

\subsubsection{Derivation of Observed Practice Effort Scores.}
Observed practice effort was derived from students’ submitted AI-agent conversation logs collected after each of the three scheduled practice sessions. Two raters independently applied a consensus-developed rubric to each submission, reaching agreement through structured discussion prior to scoring. The rubric distinguished engaged submissions — characterized by codes for substantive engagement (E) and surface engagement (SE) — from compliance-driven submissions (CE; responses that fulfilled the task requirement without evidence of genuine reasoning) and non-engaged or unusable submissions (NE).

 Each student earned up to three engagement credits (one per session), converted to a proportional score on a 0.00–1.00 scale (0.00, 0.33, 0.66, 1.00). Where strict proportional scoring did not fully capture the observed engagement pattern, raters applied qualitative notes to adjust scores within a band (e.g., 0.77, 0.88), yielding an ordinal measure of behavioral engagement distinct from self-report or time-on-task.
\subsection{Statistical Analyses }
To examine the effect of the experimental condition on learning and observed practice effort while controlling for familywise Type I error, a multivariate analysis of covariance (MANCOVA) was conducted, with posttest performance and observed practice effort entered as joint dependent variables. These two outcomes differ in scale and granularity and were not intended to be directly compared in magnitude; they were modeled jointly solely to protect against inflated Type I error across theoretically related outcomes. Experimental condition (three levels) was treated as the between-subjects factor, and baseline (pretest) performance was included as a covariate. Each dependent variable was subsequently interpreted on its own scale through follow-up univariate ANCOVAs. Multivariate significance was evaluated using Wilks'~$\Lambda$.

Statistics were based on all cases with valid data (N = 89; G1: n = 27; G2: n = 28; G3: n = 34). The homogeneity of regression slopes assumption was not statistically significant. Box’s M test was not significant, Box’s M = 11.85, F(6, 145,777.16) = 1.91, p = .076, supporting multivariate homogeneity. Levene’s tests indicated that the homogeneity-of-variance assumption was satisfied for posttest performance, F(2, 86) = 0.62, p = .539, but violated for observed practice effort, F(2, 86) = 3.58, p = .032; univariate inferences for effort were therefore interpreted with caution. Effect sizes are reported as partial $\eta^2$; standardized mean differences (Cohen’s~$d$) were computed.

\section{Results}
\subsubsection{Multivariate Effects of Condition.}
MANCOVA revealed a statistically significant multivariate effect of condition,Wilks'~$\Lambda = .664$, $F(4,168) = 9.56$, $p < .001$, partial~$\eta^2 = .185$, indicating that approximately 18.5\% of the variance in the combined outcome space was attributable to condition after adjusting for baseline. Pretest performance did not significantly predict the combined outcomes, Wilks'~$\Lambda = .998$, $F(2,84) = 0.09$, $p = .918$, partial~$\eta^2 = .002$.


\subsubsection{Univariate Effects on Observed Practice Effort.}
Follow-up ANCOVA indicated a significant effect of condition on observed practice effort, $F(2,85) = 13.58$, $p < .001$, partial~$\eta^2 = .242$. The covariate was not a significant predictor of observed practice effort, $F(1,85) = 0.03$, $p = .859$, partial~$\eta^2 \approx .000$. Adjusted marginal means showed G1 demonstrated the highest effort ($M = 0.85$, $SE = 0.05$, 95\% CI $[0.74, 0.96]$), followed by G2 ($M = 0.74$, $SE = 0.05$, $95\%$ CI $[0.64, 0.85]$) and G3 ($M = 0.49$, $SE = 0.05$, $95\%$ CI $[0.39, 0.58]$). Bonferroni-adjusted comparisons indicated that G1 and G2 did not differ significantly (p = .493), whereas both outperformed G3 (G1 vs. G3: d = 1.33, p < .001; G2 vs. G3: d = 0.83, p = .002).



\subsubsection{Univariate Effects on Posttest Performance.}
ANCOVA also revealed a significant effect of condition on posttest performance, $F(2,85) = 6.24$, $p = .003$, partial~$\eta^2 = .128$. Adjusted means indicated G1 achieved the highest posttest scores (M = 78.19, SE = 2.37), followed by G2 (M = 74.55, SE = 2.33) and G3 (M = 67.28, SE = 2.12). G1 significantly outperformed G3 (d = 0.92, p = .003); differences between G1 and G2 (p = .830) and between G2 and G3 (d = 0.57, p = .071) did not reach the adjusted significance threshold.



\section{Discussion}
This study examined whether learning gains following GenAI-enabled adaptive pretesting persist over time and how different forms of AI-supported practice influence retention and engagement. Results provide converging evidence that AI-supported practice structure plays a critical role in sustaining both effortful engagement and long-term learning after pretesting.

\subsubsection{GenAI-Enabled Adaptive Retrieval Practice Sustains Learning.}
Adaptive spaced retrieval practice (G1) produced the highest posttest performance, significantly outperforming learner-directed AI study (G3). These findings suggest that adaptive pretesting alone is insufficient to guarantee long-term retention; its benefits are best preserved when followed by structured retrieval practice that continues to challenge learners over time. This pattern aligns with evidence from the learning sciences demonstrating that retrieval practice—particularly when spaced—supports long-term retention more effectively than unguided study \cite{hopkins2016spaced,karpicke2007expanding,latimier2021meta}. Critically, this effect holds in GenAI-supported environments, where learning outcomes depend on whether AI access is structured to elicit retrieval rather than enable passive review.



\subsubsection{GenAI-Enabled Adaptive Pretesting as a Front-Loaded Catalyst.}
Adaptive pretesting appears to elevate learners’ initial understanding, but the durability of these gains depends on how subsequent learning activities are structured. When followed by retrieval-based practice—especially adaptive retrieval—pretesting contributes to sustained learning and higher-quality engagement over time. When followed only by learner-directed AI interaction, however, its benefits may attenuate.

\subsubsection{Limitations and Future Directions.}
This study has several limitations that point to directions for future research. Observed practice effort was derived from qualitative analysis of interaction logs and reflects behavioral indicators rather than internal motivational states. Future work could combine such measures with process data (e.g., timing, revision behavior) or self-report measures to provide a more comprehensive account of learner engagement. In addition, the study was conducted within a single instructional context; replication across domains and task types will be important for assessing generalizability. 


%
%
\bibliographystyle{splncs04}
\bibliography{bibliography}

\end{document}